\documentclass[onecolumn,preprintnumbers,amsmath,amssymb,12pt,nofootinbib]{revtex4}

\usepackage{graphicx}
\usepackage{dcolumn}
\usepackage{bm}
\usepackage{dsfont}
\usepackage{amssymb,amsfonts,amsmath}
\usepackage{graphicx}

\newcommand{\dd}{\mathrm{d}}

\newcommand{\ii}{\mathrm{i}}

\newcommand{\E}{\mathcal E}

\normalsize

\begin{document}

\title{\Large Fully pseudospectral time evolution and its application to $1+1$ dimensional physical problems}

\author{\bf J\"org Hennig}
\email{jhennig@maths.otago.ac.nz}
\affiliation{Department of Mathematics and Statistics,
           University of Otago,
           P.O. Box 56, Dunedin 9054, New Zealand}


\begin{abstract}
 It was recently demonstrated that time-dependent PDE problems can numerically be solved with a fully pseudospectral scheme, i.e.~using spectral expansions with respect to both spatial \emph{and} time directions (Hennig and Ansorg, 2009 \cite{Hennig2009}). This was done with the example of simple scalar wave equations in Minkowski spacetime. Here we show that the method can be used to study interesting physical problems that are described by systems of nonlinear PDEs. To this end we consider two 1+1 dimensional problems: radial oscillations of spherically symmetric Newtonian stars and time evolution of Gowdy spacetimes as particular cosmological models in general relativity.\\[1.5ex]
 \emph{Keywords}: spectral methods; nonlinear PDEs; stellar oscillations; general relativity
\end{abstract}


\maketitle
\section{Introduction \label{sec:Intro}}

Many dynamical physical phenomena can be mathematically modelled in terms of (hyperbolic) partial differential equations (PDEs). Hence, it is of fundamental interest to study properties of PDEs and, of course, to construct their solutions. In the special case of \emph{linear} PDEs, quite comprehensive theory is available and it is possible to find exact solutions in many cases. On the other hand, there are large gaps in the mathematical theory of \emph{nonlinear} equations and exact solutions are rare exceptions. However, many (if not most) interesting processes in nature can be described by nonlinear PDEs, so that it is desirable to have methods for the construction of at least approximate solutions in these important cases. One approach could be the application of analytical approximation techniques, and another possibility are \emph{numerical methods}. In this article, we focus on this second alternative.

There is a large variety of numerical methods available that can be used to solve nonlinear PDEs. If one is interested in highly accurate solutions (in order to be as close to an exact solution as possible), one particularly useful approach is (pseudo-) spectral methods. In the case of \emph{dynamical} (time-dependent) problems, traditionally a combination of spectral methods (with respect to the spatial directions) and finite difference methods (in time direction) has been used. We refer the reader to \cite{Grandclement} for a comprehensive overview of spectral methods in general relativity as a particular and interesting field of application. Until recently, however, there was almost no experience with spectral expansions with respect to space \emph{and} time\footnote{To the best of our knowledge there is only one article that studies \emph{parabolic} equations by means of spectral methods in space and time \cite{Ierley}.}. This changed when it was shown in \cite{Hennig2009} that a fully pseudospectral 
scheme can indeed be successfully applied to hyperbolic PDEs. With the example of simple 1+1 dimensional scalar wave equations in Minkowski spacetime, it was demonstrated in \cite{Hennig2009} that highly accurate numerical solutions can be obtained in this way.  A first application of this method is given in \cite{Ansorg2011}, where the general relativistic field equations have been solved in the interior region of axisymmetric and stationary black holes. 

In this paper, we present two new applications of the fully pseudospectral scheme  to 1+1 dimensional dynamical problems in the context of Newtonian and Einsteinian gravity. Our intention is to demonstrate, with interesting and nontrivial examples, that this approach is a very promising method, which seems to be applicable to a large class of nonlinear PDE problems. In particular, we show that stable and highly accurate long-term evolutions are possible.

This paper is organized as follows. We start by summarizing the basic idea of the fully pseudospectral scheme in Sec.~\ref{sec:method}. In Sec.~\ref{sec:Newton}, we consider radial oscillations of spherically symmetric stars within Newton's theory of gravitation as a first application of the method. Afterwards, the time evolution of particular general relativistic cosmological models (Gowdy spacetimes with spatial $S^3$ topology) is studied as a second application in Sec.~\ref{sec:gowdy}. Finally, we conclude with a discussion of our results in Sec.~\ref{sec:discussion}.

\section{Description of the numerical method \label{sec:method}}

In the following we summarize the ideas of the fully pseudospectral scheme. For more details we refer to \cite{Hennig2009}. 
The underlying numerical method is a generalization of the method by Ansorg et al.  for solving elliptic PDEs \cite{Ansorg2002} to the case of hyperbolic equations.


Our goal is to construct approximations to functions $f_1, f_2,\dots,f_m$ (depending on a spatial coordinate, say $x$, and on time $t$) that solve a system of $m$ coupled, nonlinear PDEs subject to some initial conditions. Depending on the order of the equations, the initial conditions could be prescribed function values and derivatives up to some order at some given instant of time $t=t_\textrm{min}$. The numerical solution should then be calculated in a time interval $[t_\textrm{min},t_\textrm{max}]$, i.e.~we assume that we are interested in a compact time domain. Moreover, we assume that also $x$ takes on values in a bounded domain only\footnote{In the following examples, this will automatically be the case as a consequence of the physical setting (compact interior region of an oscillating star/compact cosmological model). However, one could also start from a problem formulated on an unbounded spatial (or/and time) domain and perform a compactification. An example for this precedure is the conformal 
compactification of Minkowski spacetime discussed in~\cite{Hennig2009}.}. In addition to initial conditions, one generally expects to have some boundary conditions at the boundaries of the spatial domain. These conditions might either come from some physical assumptions or from a particular degeneracy of the equations at the boundaries so that regular solutions are only possible if appropriate boundary conditions are satisfied.

Our pseudospectral scheme starts by mapping the physical domain of the PDE problem to one or several unit squares by means of a coordinate transformation to spectral coordinates $\sigma, \tau\in[0,1]\times[0,1]$ in each domain. In some cases this transformation is a simple rescaling of the physical variables $x$ and $t$ with constant factors.
However, in other cases the original domain might not even be known from the beginning, i.e.~one may also be interested in \emph{free boundary value problems}. (An example is provided by the stellar pulsations to be considered in Sec.~\ref{sec:Newton}, where the interior of the star is a sphere of radius $R(t)$. However, the function $R=R(t)$ can only be determined together with the unknown fluid variables while the hydrodynamic equations are solved and is, therefore, not known \emph{a priori}.) As a consequence, the coordinate transformations may involve unknown functions which have to be determined from some additional conditions together with the other unknown functions $f_i$, $i=1,\dots,m$.

In a next step, we need to ensure that the initial conditions at $\tau=0$ (corresponding to $t=t_\textrm{min}$) are satisfied. For that purpose, we write the functions $f_i(\sigma,\tau)$ in terms of functions $\tilde f_i(\sigma,\tau)$ in such a way that the initial conditions are satisfied automatically. As an example, consider an initial value problem where the function values at $\tau=0$ are given, $f_i(\sigma,0)=g_i(\sigma)$. Then introduce new unknowns $\tilde f_i$ via
\begin{equation}\label{eq:incon}
 f_i(\sigma,\tau)=g_i(\sigma)+\tau \tilde f_i(\sigma,\tau).
\end{equation}
If also derivatives of $f_i$ need to be prescribed at the initial surface, then one can choose a similar expansion including higher order terms in $\tau$, see \cite{Hennig2009}.

Once the mathematical problem is reduced to a PDE problem on unit squares and we have taken care of the initial conditions, the main idea of a pseudospectral method is to approximate the unknown functions $\tilde f_i$ in terms of finite linear combinations of some set of basis functions. In our scheme we choose Chebyshev polynomials $T_k$ and use an approximation of the form
\begin{equation}\label{eq:approx}
 \tilde f_i(\sigma,\tau)\approx\sum\limits_{j=0}^{N_\sigma}\sum_{k=0}^{N_\tau}
        c_{ijk}T_j(2\sigma-1)T_k(2\tau-1),\quad i=1,\dots,m,
\end{equation}
for given spectral resolutions (number of polynomials) $n_\sigma \equiv N_\sigma+1$, $n_\tau \equiv N_\tau+1$. Note that it is a simple task to find approximations for the spatial and time derivatives of $\tilde f_i$ once we have the approximation \eqref{eq:approx}. 

Now we have to determine the Chebyshev coefficients $c_{ijk}$. To this end, we introduce a suitable set of collocation points. Here, we choose \emph{Gauss-Lobatto} nodes\footnote{It is well-known that a nearly optimal polynomial approximation is achieved for \emph{Gauss-Chebyshev} collocation points (Chebyshev roots), see, e.g., \cite{Boyd}. The \emph{Gauss-Lobatto} points used here (Chebyshev extrema plus boundary points) lead to slightly less accurate approximations. However, they have the advantage to include gridpoints at the boundaries of our unit squares, which is useful for implementing boundary conditions.}, which are defined by
\begin{equation}
 \sigma_j=\sin^2\left(\frac{\pi j}{2N_\sigma}\right),\quad
 \tau_k=\sin^2\left(\frac{\pi k}{2N_\tau}\right),\quad
 j=0,\dots,N_\sigma,\quad
 k=0,\dots,N_\tau.
\end{equation}
Then we impose the condition that our approximations to the functions $f_i$, together with the corresponding approximations to their derivatives, satisfy the PDEs (plus the boundary conditions, if applicable) exactly at these points. This provides us with a large algebraic system of equations for the Chebyshev coefficients $c_{ijk}$ or, equivalently, for the function values $\tilde f_i(\sigma_j,\tau_k)$ --- and this is the formulation that we actually use. 

Finally, starting from some initial guess, we solve this system iteratively with the \emph{Newton-Raphson} method. This part of the algorithm makes the method computationally expensive, since inversion of relatively large matrices is required. However, in the present case of 1+1 dimensional problems and for a moderate number of unknown functions $f_i$ and gridpoints ($n_\sigma,n_\tau \sim 30$) this is no problem and the method will be reasonably fast.

There are quite different ways of obtaining an initial guess, for which the Newton-Raphson method will converge. In some cases it is possible to start from the trivial solution. Sometimes, one can introduce a parameter, say $\alpha$, into the problem, such that the solution is known for $\alpha=0$ and the relevant solution is obtained for $\alpha=1$. Then, starting from the case with $\alpha=0$, one can go to $\alpha=1$ in a few steps and always use the solution from the previous step as an initial guess for the next one. This procedure is used for the calculation of oscillating stars in Sec.~\ref{sec:Newton}, where we can start from initial data corresponding to a star in equilibrium. The finally interesting star (with initial data that significantly deviate from equilibrium data) is then reached in about one to three steps. It is also possible to start from an analytically obtained approximate solution, for example by using Taylor expansions about the initial hypersurface. This method is used in Sec.~\ref{sec:gowdy} for the calculation of cosmological models. Finally, one could also solve the problem with, e.g., a finite difference method first and interpolate the resulting numerical values onto the spectral grid to obtain an initial guess.

The described pseudospectral method is a \emph{highly implicit} numerical scheme\footnote{With the term ``\emph{highly} implicit'' we want to emphasize that the numerical values at \emph{all} (instead of, perhaps, only two or three) time slices are coupled in our method.}, in which the solution at $n_\sigma$ points in spatial direction times $n_\tau$ points in time direction is obtained simultaneously. This is in contrast to the approach in most other numerical PDE methods, where the solution is usually calculated at one time slice after the other. However, this implicit character plays a crucial role for obtaining highly accurate solutions and is, therefore, an essential part of the fully pseudospectral scheme. 

\section{Newtonian stellar oscillations\label{sec:Newton}}

We come now to the main part of this paper and study applications of the fully pseudospectral scheme. As a first example, we consider nonlinear radial stellar oscillations within Newton's theory of gravitation.

\subsection{The physical model}

\emph{Variable stars} are a class of astrophysical objects whose luminosity (or, more precisely, apparent magnitude as seen from Earth) changes over time. 
This fascinating phenomenon has already been intensively studied for a long time, both observationally and theoretically. Among the results is the famous 
period-luminosity relation \cite{Leavitt} for classical Cepheids.
The idea that, for some types of variable stars, the luminosity changes are caused by \emph{stellar pulsations} dates back almost 100 years to work by Plummer \cite{Plummer} and Shapley \cite{Shapley}\footnote{Interesting details on the history of the theory of stellar pulsations can be found in \cite{Gautschy}.}. Nowadays, stellar pulsations are used as an accurate tool to study many aspects of stellar physics. In particular, the numerical simulation of stellar oscillations is an important element of the investigation of properties of variable stars.
In recent times, oscillations have also attracted much attention in the context of compact astrophysical objects such as \emph{neutron stars}, in particular in the field of gravitational wave physics. 
For an overview over the Newtonian description of stellar pulsations we refer to \cite{Cox}. An interesting recent numerical study (based on a finite difference method) of general relativistic radial pulsations can be found in \cite{Gabler}.

Since our main interest is in the numerical method, we neglect the effects of general relativity and study here the simpler equations of Newtonian physics. To this end, we model an oscillating star as a fluid ball of variable radius $R=R(t)$, described by the following equations,
\begin{eqnarray}
 &\bigtriangleup U=4\pi G\rho & \quad\textrm{(Poisson equation)},\\
 &\rho\frac{\dd\mathbf{v}}{\dd t}
  \equiv \rho\left[\frac{\partial\mathbf{v}}{\partial t}+(\mathbf{v}\cdot\nabla)\mathbf{v}\right]
   =-\nabla p-\rho\nabla U & \quad\textrm{(Euler equation)},\\
 &\frac{\partial\rho}{\partial t}+\nabla\cdot(\rho\mathbf{v})=0
  & \quad\textrm{(Continuity equation)}\label{eq:con},
\end{eqnarray}
where $U$, $\rho$, $p$, $\mathbf{v}$ and $G$ denote the gravitational potential, the mass density, the pressure, the velocity field and the gravitational constant, respectively. In addition, we choose a polytropic equation of state
\begin{equation}\label{eq:eos}
 p=K\rho^\gamma,\quad \gamma=1+\frac{1}{n},
\end{equation}
with polytropic index $n=1$ ($\gamma=2$), where $K$ is the polytropic constant.

In order to obtain a $1+1$ dimensional problem, we assume spherical symmetry, where all functions depend only on a radial coordinate $r$ and on time $t$. This results in the simplified system of equations
\begin{eqnarray}
 U_{,rr}+\frac{2}{r}U_{,r}=4\pi G\rho,\label{eq:U}\\
 \rho(v_{,t}+v v_{,r})=-p_{,r}-\rho U_{,r}\label{eq:v},\\
 \rho_{,t}+(\rho v)_{,r}+\frac{2}{r}\rho v=0\label{eq:rho},\label{eq:con1}
\end{eqnarray}
where $v$ is the radial velocity and the subscripts denote derivatives with respect to the indicated variables. Finally, after eliminating the pressure with \eqref{eq:eos}, we arrive at a system of three coupled PDEs for the three functions $U$, $\rho$ and $v$. Since the equations \eqref{eq:v} and \eqref{eq:rho} are first order in time, we can prescribe the initial values for $v$ and $\rho$. But of course we cannot choose an initial potential $U$, since $U$ is already determined by $\rho$ from \eqref{eq:U}.

Regular functions $U$ and $\rho$ inside the star have the property of being even in $r$, whereas $v$ is an odd function of $r$. Hence, if we replace $v$ with 
\begin{equation}
 w(r,t):=\frac{v(r,t)}{r},
\end{equation}
then we obtain a set of three even functions, $U$, $\rho$, $w$, which can be considered as functions of $r^2$ and $t$. This will be used when we introduce adapted spectral coordinates in the next subsection.

For an initial configuration at time $t=0$ we choose a slightly disturbed equilibrium star. In equilibrium (no time-dependence), the above equations for a polytropic star can be reduced to the Lane-Emden equation \cite{Emden}. For $n=1$, the exact solution is known and we obtain the corresponding equilibrium solution
\begin{equation}
 \rho_\textrm{eq}(r)=\rho_\textrm{c}\, \mathrm{sinc}\left(\pi \frac{r}{r_0}\right),\quad
 U_\textrm{eq}(r)=-2K\rho_\textrm{c}\left[1+\mathrm{sinc}\left(\pi\frac{r}{r_0}\right)\right],\quad w_\textrm{eq}(r)=0,
\end{equation}
where $\rho_c$ is the central mass density, $r_0=\sqrt{\frac{\pi K}{2G}}$ is the equilibrium radius and $\mathrm{sinc}\, x=\frac{\sin x}{x}$. One possibility of disturbing this equilibrium is to consider a non-vanishing initial velocity field, i.e.~we can start from the initial data
\begin{equation}\label{eq:ic}
 \rho(r,0)=\rho_\textrm{eq}(r),\quad
 U(r,0)=U_\textrm{eq}(r),\quad
 w(r,0)=w_0(r)\not\equiv 0
\end{equation}
for some function $w_0=w_0(r)$.

Finally, we note that the exterior vacuum region around an oscillating star is trivially described by the potential $U=-GM/r$, where $M$ is the (constant) mass of the star. Hence, we can restrict our attention to the interior region. The boundary between both regions (i.e.~the rim of the star) is characterized by vanishing pressure, $p=0$, and, therefore, by $\rho=0$ for our equation of state \eqref{eq:eos}.

\subsection{Numerical details}

In order to study the time evolution of a star over a longer period of time (covering several oscillation periods), we divide the time interval $[0,t_\textrm{max}]$ into $N$ subintervals 
\begin{equation}
 [t_0,t_1], [t_1,t_2], \dots, [t_{N-1},t_N] \quad\textrm{with}\quad
 t_i=i\Delta t, \quad \Delta t:=\frac{t_\textrm{max}}{N}.
\end{equation}
In this way we can ensure that the unknown functions do not oscillate too strongly in each domain. Otherwise, Chebyshev polynomials of much higher degree would be required for accurate approximations.

For the first time domain, initial conditions are given at $t=0$ as described in the previous subsection (perturbations of the Lane-Emden equilibrium solution, cf. Eq.~\eqref{eq:ic}). When the functions in the first domain have been computed, new initial data are read off for the second domain, and so on. Hence, the time domains are evaluated step by step, one after the other. (In a future version of the code, a simultaneous treatment of several domains might be considered to further increase the numerical accuracy.)

Spectral coordinates $(\sigma,\tau)\in[0,1]\times[0,1]$ in each domain $i=0,\dots,N-1$ are now introduced via
\begin{equation}
 r=\sqrt{\sigma} R(t),\quad t=t_i+\tau\Delta t,
\end{equation}
where the square root takes into account that all the unknowns are regular functions of $r^2$ and $t$ as discussed above. This transformation contains the unknown radius $R=R(t)$, which can be determined from the surface condition $p(r=R(t),t)=0$. 

In terms of these new coordinates, the fluid equations \eqref{eq:U}-\eqref{eq:con1} take the following form,
\begin{eqnarray}
 & 2\sigma U_{,\sigma\sigma}+3U_{,\sigma}=2\pi GR^2\rho,\label{ES1}\\
 & \frac{R}{\Delta t}w_{,\tau}+2\sigma\left(Rw-\frac{R_{,\tau}}{\Delta t}\right) w_{,\sigma}
   +Rw^2+\frac{2}{R}(2K\rho_{,\sigma}+U_{,\sigma})=0,\\
 & \frac{1}{\Delta t}\rho_{,\tau}+2\sigma\left(w-\frac{R_{,\tau}}{R\Delta t}\right)
    \rho_{,\sigma}+\rho(2\sigma w_{,\sigma}+3w)=0.
\end{eqnarray}
Note that these equations contain only spatial derivatives of $U$, but not $U$ itself. Hence, $U$ is determined only up to an additive function of $\tau$ (or, equivalently, of $t$). This is physically reasonable, since only the potential difference (work) or the gradient (force) are measurable quantities. Here, we fix this degree of freedom in the usual way by choosing $U\to 0$ at infinity. As a consequence, we have the typical $1/r$-potential of spherical symmetry in the exterior region. Since every $1/r$-potential satisfies $rU_{,r}+U=0$, this condition must also hold for the interior potential at the boundary $r=R$, as a consequence of a continuous and differentiable transition at this boundary. Consequently, we impose this equation instead of Eq.~\eqref{ES1} at the rim of the star. In terms of the spectral coordinates, this boundary condition becomes
\begin{equation}
 2\sigma U_{,\sigma}+U=0.
\end{equation}
Alternatively, one could also prescribe the values of $U$ at some timelike curve, for example on the symmetry axis, e.g.\ by setting $U$ to zero there. But this would introduce an ``artificial'' time dependence of the exterior potential.

The initial conditions for $\rho$ and $w$ are implemented by replacing these functions in terms of functions $\tilde\rho$ and $\tilde w$ as illustrated in Eq.~\eqref{eq:incon}. Therefore, we finally arrive at the set of unknown quantities $U(\sigma,\tau),\tilde\rho(\sigma,\tau),\tilde w(\sigma,\tau),R(\tau)$, which can be calculated by solving the three fluid equations together with the boundary condition for $U$ plus the extra condition of vanishing surface pressure.

\subsection{Numerical results}

\begin{figure}\centering
 \hspace{.2cm}(a)\hspace{7.95cm}(b)\hfill\mbox{}\\
 \includegraphics[scale=0.63]{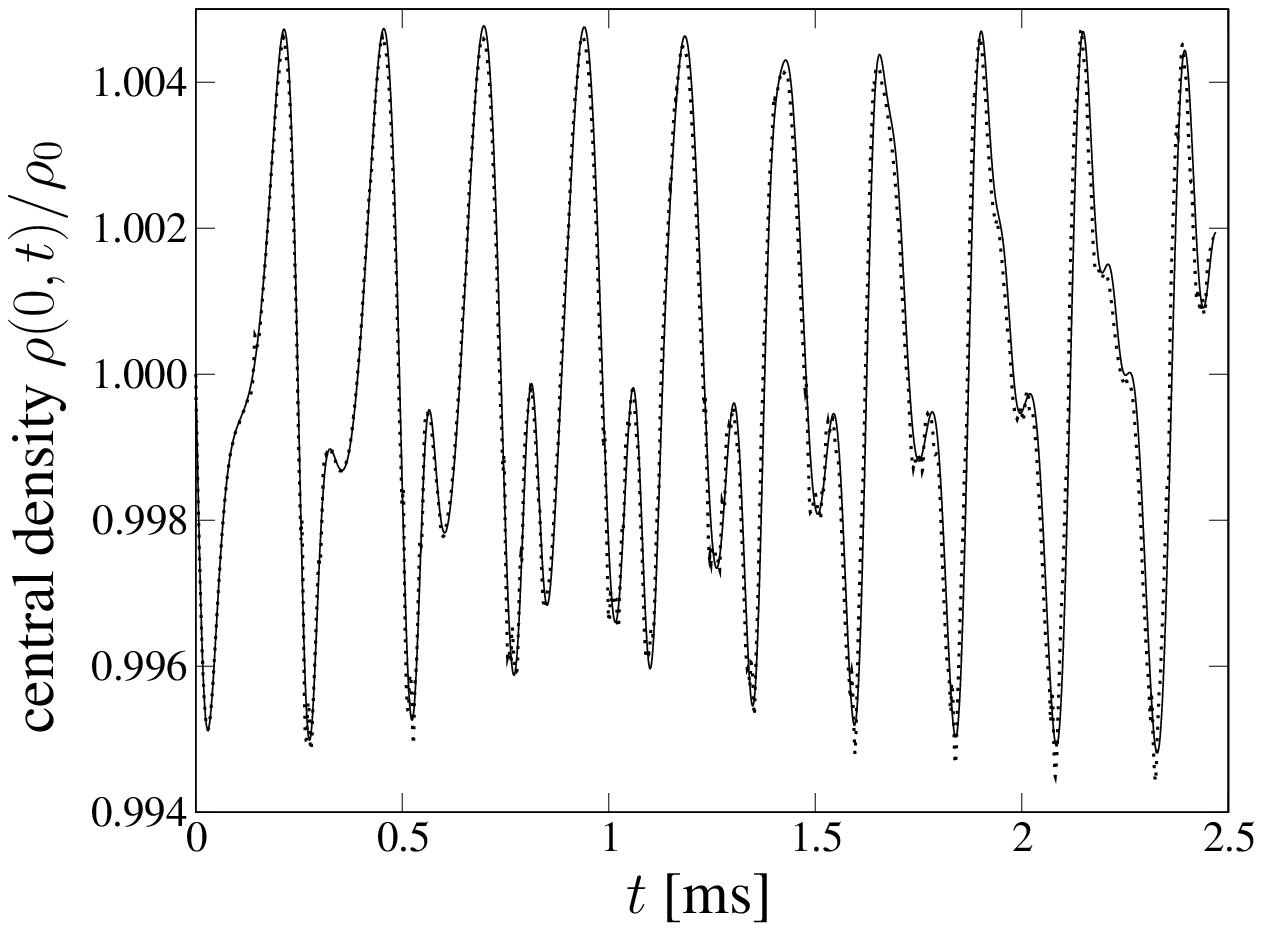}\quad\,
 \includegraphics[scale=0.63]{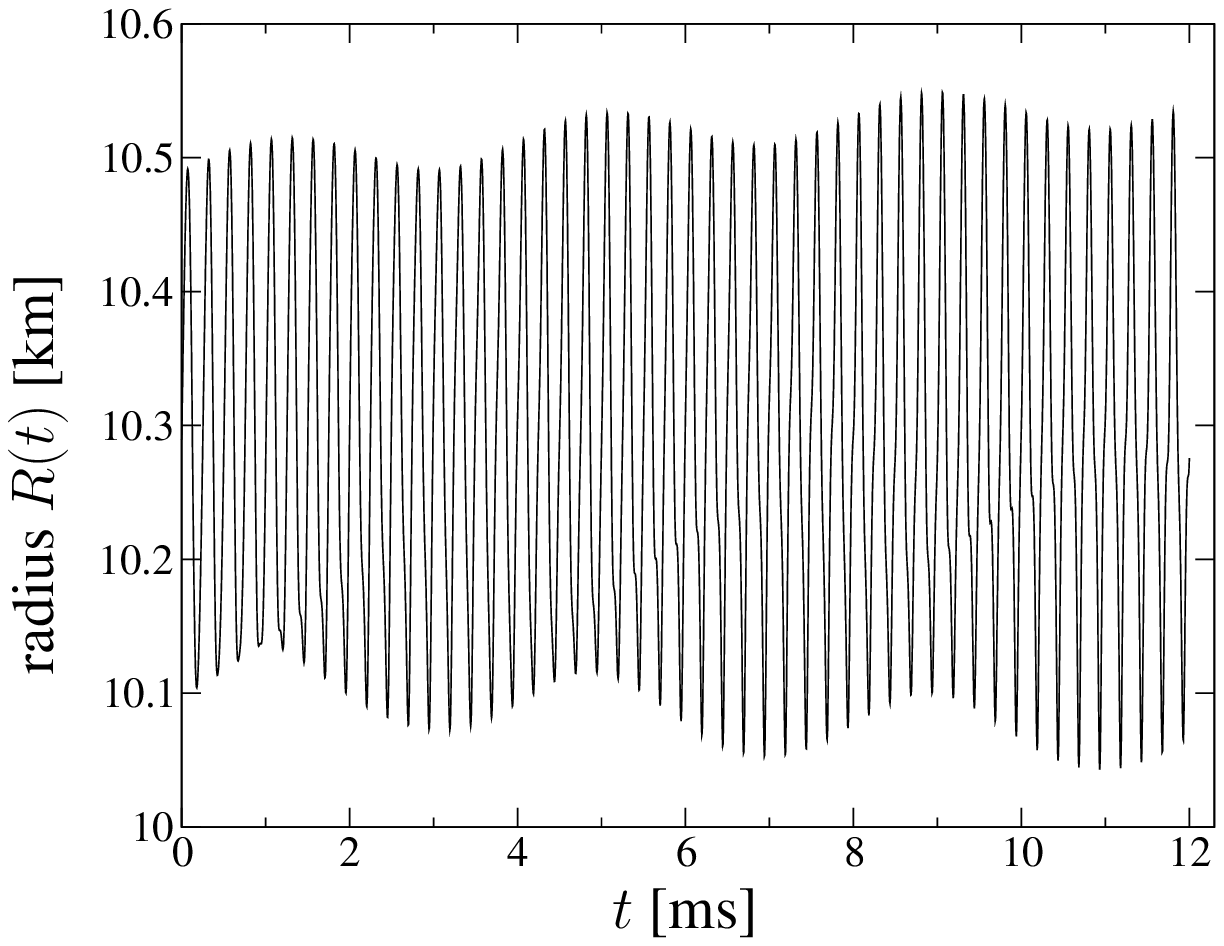}
 \caption{\label{fig:NS}
 Two examples of spectral evolutions (solid curves) with different initial data. The dotted curve in the left panel shows for comparison the result as obtained with a finite difference code.\\
 (a): central density for $v(r,0)=300\mathrm{\frac{km}{s}}\times\sin(\pi r/r_0)$, $40$ time domains with resolution $16\times 18$\\
 (b): radius for $v(r,0)= 4116\mathrm{\frac{km}{s}}\times r/r_0$, 150 time domains with resolution $18\times 20$.}
\end{figure}

In the following, we consider examples of Newtonian stars as simple models for \emph{neutron stars}, i.e.~we choose parameters leading to stars with radii about $10\,\mathrm{km}$ and masses about $1.4\,M_\odot$. To this end, we start from the equilibrium density $\rho(r,0)=\rho_\textrm{eq}(r)$
with initial central density $\rho_\textrm{c}=1.9891\times 10^{18}\frac{\textrm{kg}}{\textrm{m}^3}$. As discussed above, the equilibrium is then distorted by adding an initial velocity field $w(r,0)$. Furthermore, we choose a polytropic constant of $K=4.5\times10^{-3}\,\mathrm{\frac{m^5}{kg\,s^2}}$ so that a star in equilibrium would have a radius of $r_0=10.29\,$km. Only in the first calculation (Fig.~\ref{fig:NS}a) we use a slightly different value, namely $K=4.25\times10^{-3}\,\mathrm{\frac{m^5}{kg\,s^2}}$.

In a first numerical example, a star is evolved for $2.5\,$ms (corresponding to 10 fundamental oscillations), see Fig.~\ref{fig:NS}a, and the result is compared to data obtained with a finite difference code\footnote{I am grateful to Ernazar Abdikamalov for providing me with the finite difference result.}. The diagram shows that both curves agree well, which is a first confirmation that the pseudospectral method produces correct results.

As second example, we evolve a star with different initial conditions for $12\,$ms (48 fundamental oscillations). The result is shown in Fig.~\ref{fig:NS}b. Besides the fundamental oscillation (with frequency\footnote{The numerical values for the frequencies of the oscillation modes have been obtained from a linear perturbation analysis with a pseudospectral code. For details about linear perturbations of polytropic stars, see, e.g., \cite{Cox} or \cite{MTW}.} $f_0=4.009596\,$kHz, corresponding to a period of $0.249402\,$ms) one can see an amplitude modulation with lower frequency. This happens  since the frequencies of different oscillation modes are not precisely multiples of the fundamental frequency. As a consequence, there is a beat with frequency $f_1-2f_0=0.25\,$Hz (period of $3.99\,$ms), where $f_1=8.269880\,$kHz is the frequency of the first overtone. In addition, the amplitude increases due to another beat with frequency $f_2-3f_0=0.035\,$kHz (period of $28.2\,$ms), where $f_2=12.064215\,$kHz.

\subsection{Convergence}

As illustrated with the numerical examples in the previous subsection (in particular, by comparing with the finite difference result), a fully pseudospectral scheme is able to produce correct time evolutions of oscillating stars. However, the main reason for using a spectral method is, of course, that one wants to produce \emph{very} accurate results. Therefore, it is desirable to test convergence properties of the code and to measure the accuracy quantitatively.

To this end, we look at two different quantities and their accuracy. In a first step, we may calculate the mass $M$ of the star as a function of time,
\begin{equation}
 M(t)=4\pi\int\limits_0^{R(t)}\rho(r,t)r^2\dd r.
\end{equation}
The integral in the latter formula can be approximated by using the Chebyshev representation of $\rho(r,t)$ with the numerically calculated coefficients and integrating the Chebyshev polynomials exactly. As a consequence of the continuity equation \eqref{eq:con} [or \eqref{eq:con1}], $M$ must, of course, be independent of $t$. On the other hand, numerical errors will introduce some artificial time dependence of $M$. This effect can be used as a measure for the numerical accuracy. 

\begin{figure}\centering
 \hspace{.2cm}(a)\hspace{7.95cm}(b)\hfill\mbox{}\\
 \includegraphics[scale=0.63]{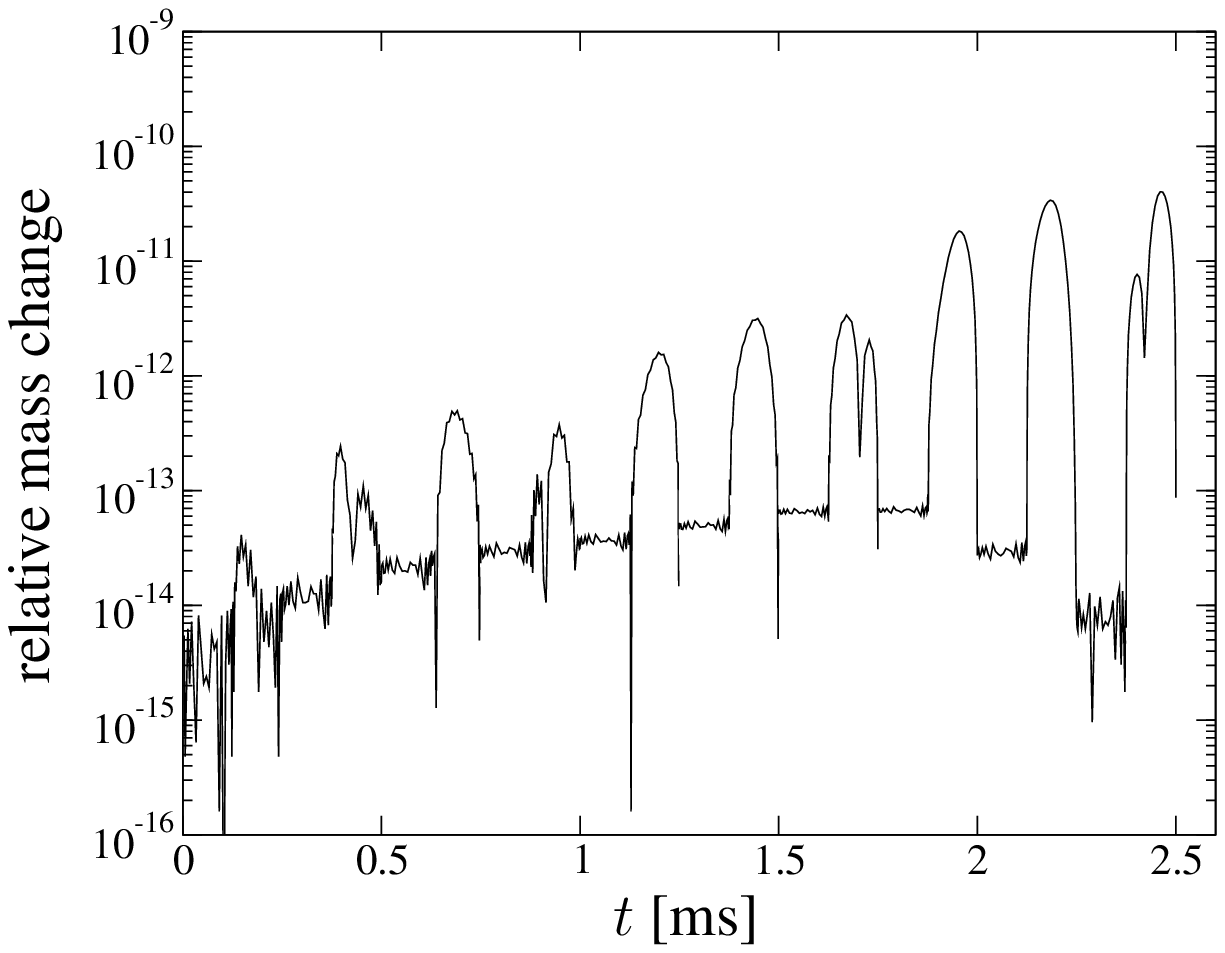}\quad\,
 \includegraphics[scale=0.63]{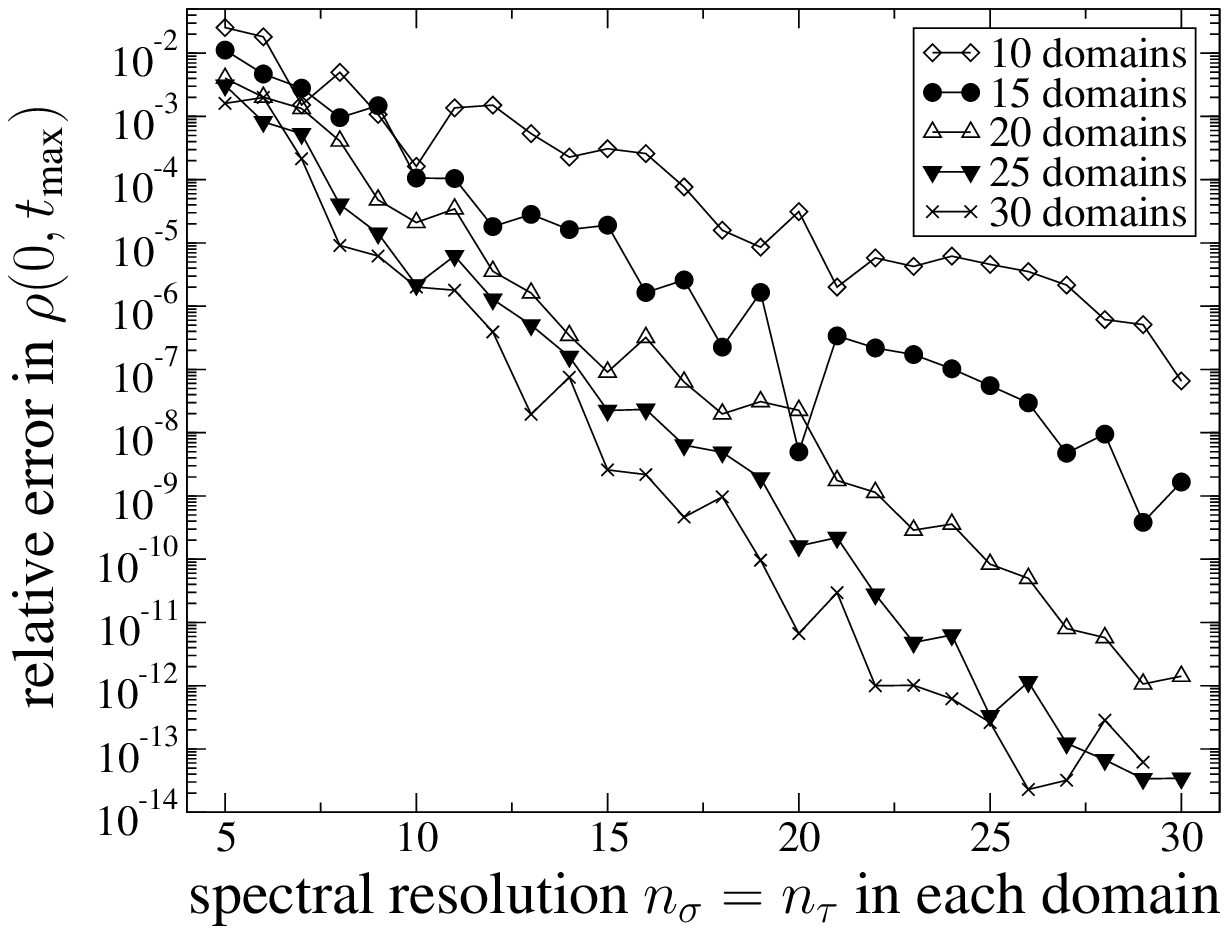}
 \caption{\label{fig:NSCon}
 Test of the numerical accuracy: A star with initial velocity profile $v(r,0)= 3087\mathrm{\frac{km}{s}}\times r/r_0$ is evolved for $2.5$ ms (10 fundamental oscillations). The figure shows:\\
 (a) The relative mass change $|[m(t)-m(0)]/m(0)|$ for 25 time domains with a resolution of $25\times 25$.\\
 (b) The relative error in the final central mass density $\rho(0,t_\textrm{max})$ for different numbers of time domains and spectral resolutions as estimated by comparing the values of $\rho(0,t_\textrm{max})$ with the value obtained for 30 domains with resolution $30\times30$.}
\end{figure}

The relative mass change during ten oscillations of a star is plotted in Fig.~\ref{fig:NSCon}a. As one might have expected, the average error is growing with time. Finally, the error reaches values between $10^{-11}$ and $10^{-10}$ at the end of the simulation, which shows that the numerical result is indeed very accurate.

Next, we investigate how the error depends on the spectral resolution $n_\sigma$, $n_\tau$ and on the number of time domains. For that purpose, we look at the numerical values for the central mass density at $t=t_\textrm{max}$ as obtained for different resolutions. These values may then be compared to the most accurate value (corresponding to the highest resolution), which can be considered to be more or less exact. In this way, we obtain a measure for the relative error in the central density at the end of the simulations (where the largest errors are expected). The result can be found in Fig.~\ref{fig:NSCon}b. Obviously, the error trends downwards for increasing spectral resolution, if the number of time domains is fixed. On the other hand, the error for given resolution in each domain decreases, if the number of time domains is increased. However, once the error has reached values in the order of $10^{-12}$ to $10^{-13}$ (close to machine accuracy for our double precision code), it cannot be reduced any 
further by considering more domains or higher resolution. 

This demonstrates that \emph{highly accurate solutions can be obtained} for moderate values of the resolution and number of time domains. Moreover, the error curves in Fig.~\ref{fig:NSCon}b are approximately linear in a logarithmic plot, i.e.~we observe \emph{exponential convergence}.

\subsection{Eigenmodes and nonlinear effects}

In order to study effects coming from the nonlinearities in the equations for stellar pulsations, we look at the differences between oscillations with small amplitudes and oscillations with large amplitudes.

If we choose initial conditions close to equilibrium (i.e.~with a small initial velocity field), then the stellar pulsations can be treated as linear perturbations of stars in equilibrium, cf. \cite{Cox,MTW}. As a result, we can expect that the time dependent quantities will undergo harmonic oscillations with particular frequencies (corresponding to the eigenmodes of the linearized equations). It should also be possible to reproduce this effect with our fully nonlinear code, provided the initial perturbations are small enough so that nonlinear terms are irrelevant. 
This can be demonstrated by choosing eigenfunctions of the linear oscillation modes with small amplitude as initial velocity field. 

\begin{figure}\centering
 \hspace{.2cm}(a)\hspace{7.95cm}(b)\hfill\mbox{}\\
 \includegraphics[scale=0.628]{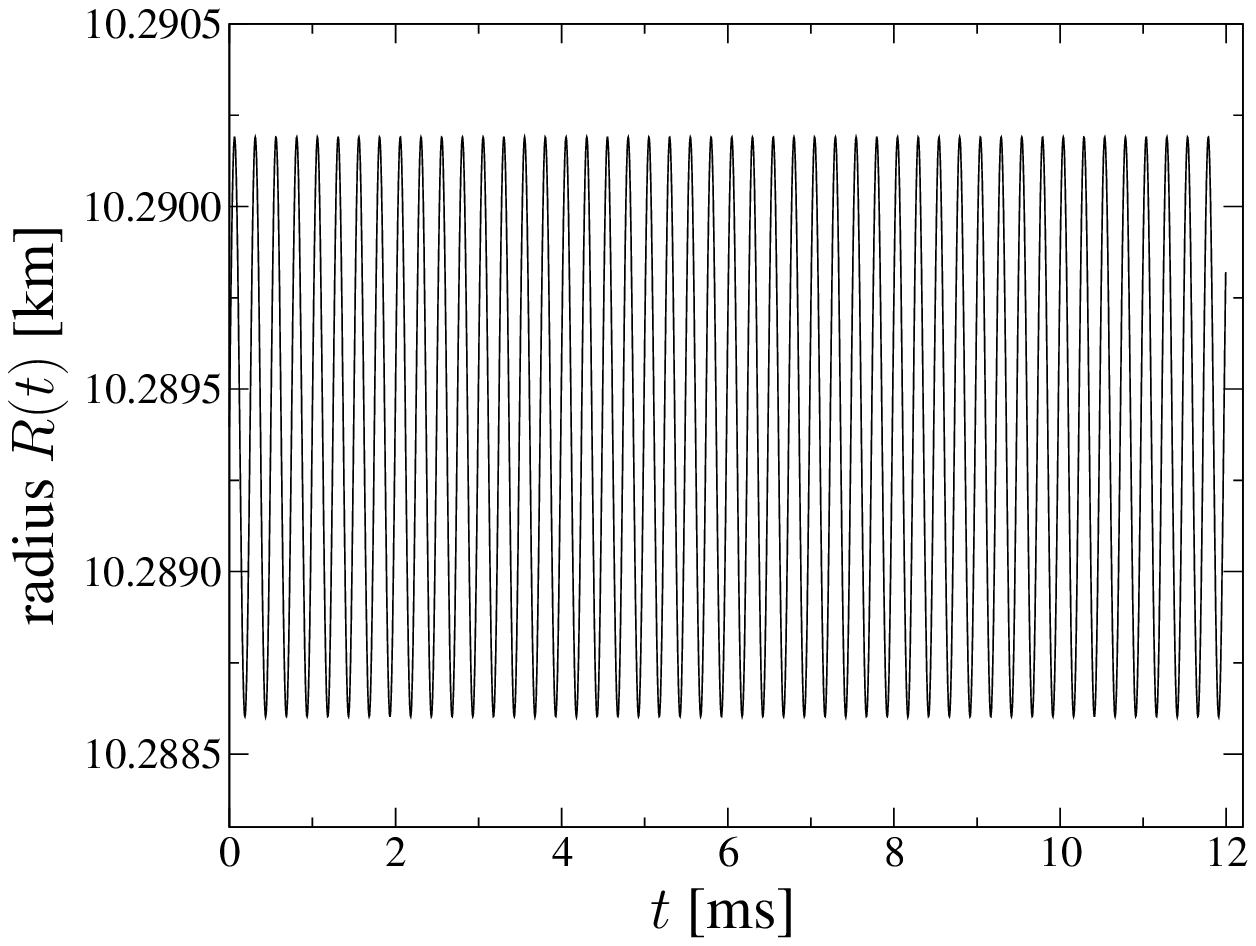}\quad\,
 \includegraphics[scale=0.628]{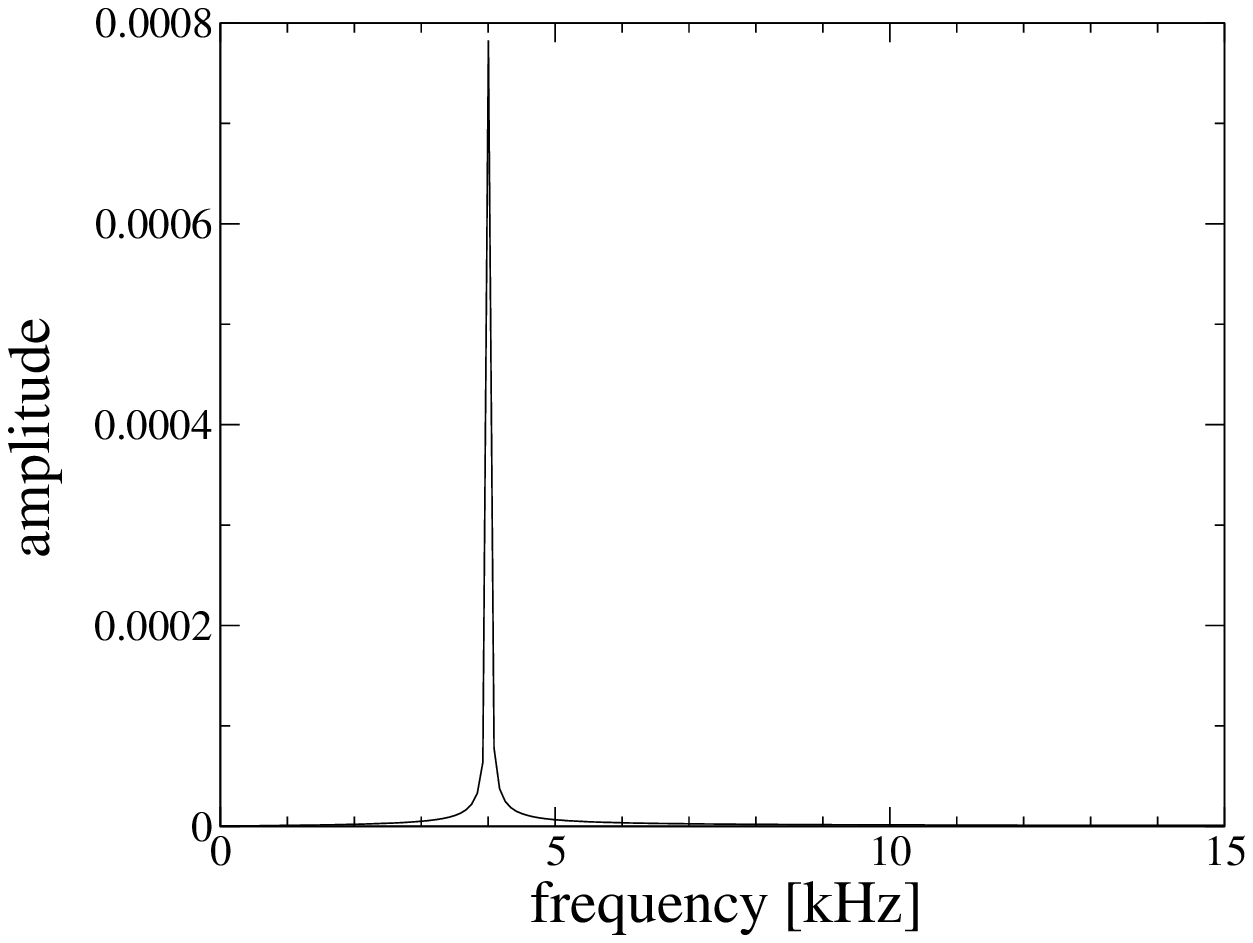}
 \caption{\label{fig:mode0}
 Fundamental mode with maximal initial velocity $v(r_0,0)=20\mathrm{\frac{km}{s}}$. The plots show (a) the radius $R(t)$ and (b) the Fourier spectrum.}
\end{figure}

\begin{figure}\centering
 \hspace{.2cm}(a)\hspace{7.95cm}(b)\hfill\mbox{}\\
 \includegraphics[scale=0.628]{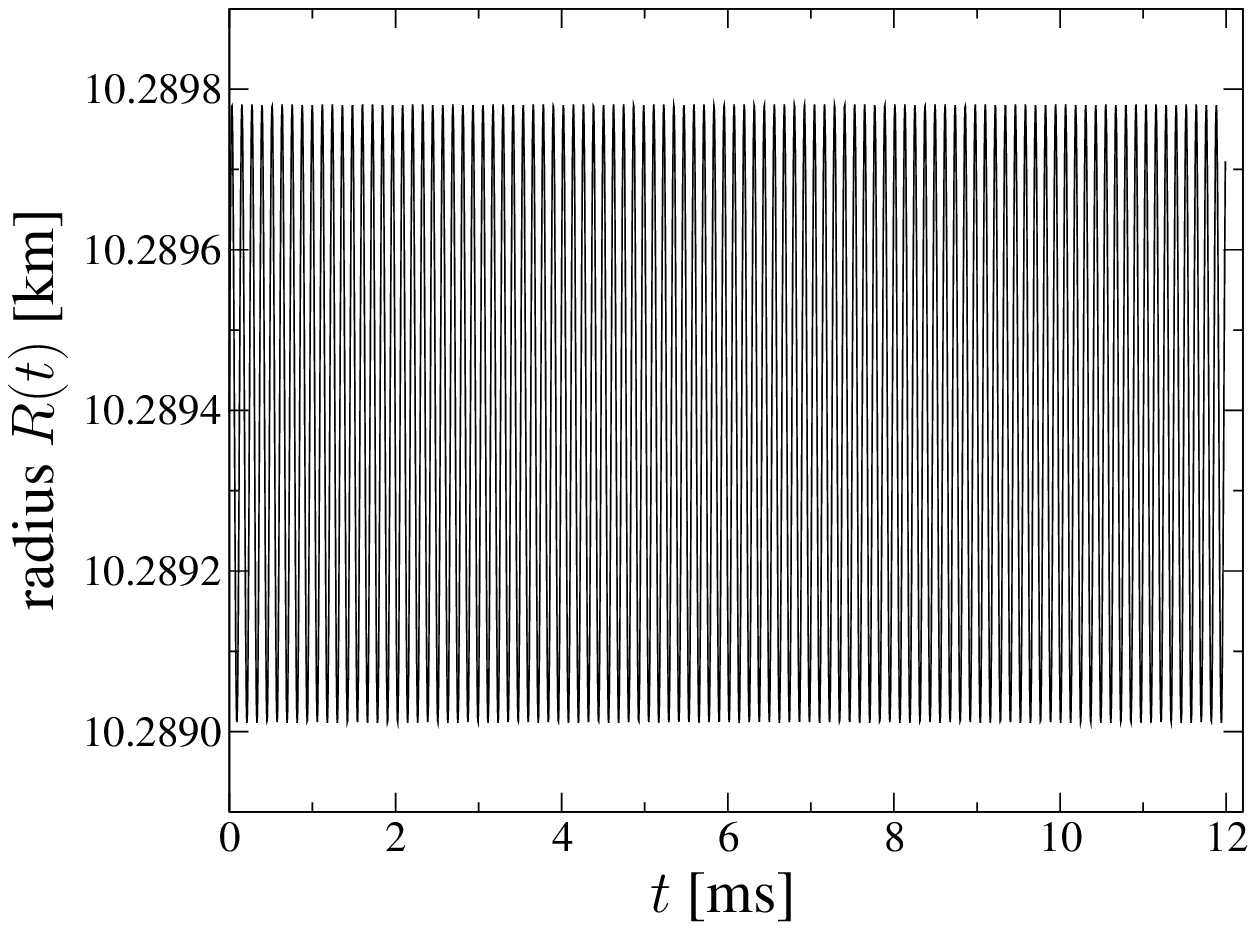}\quad\,
 \includegraphics[scale=0.628]{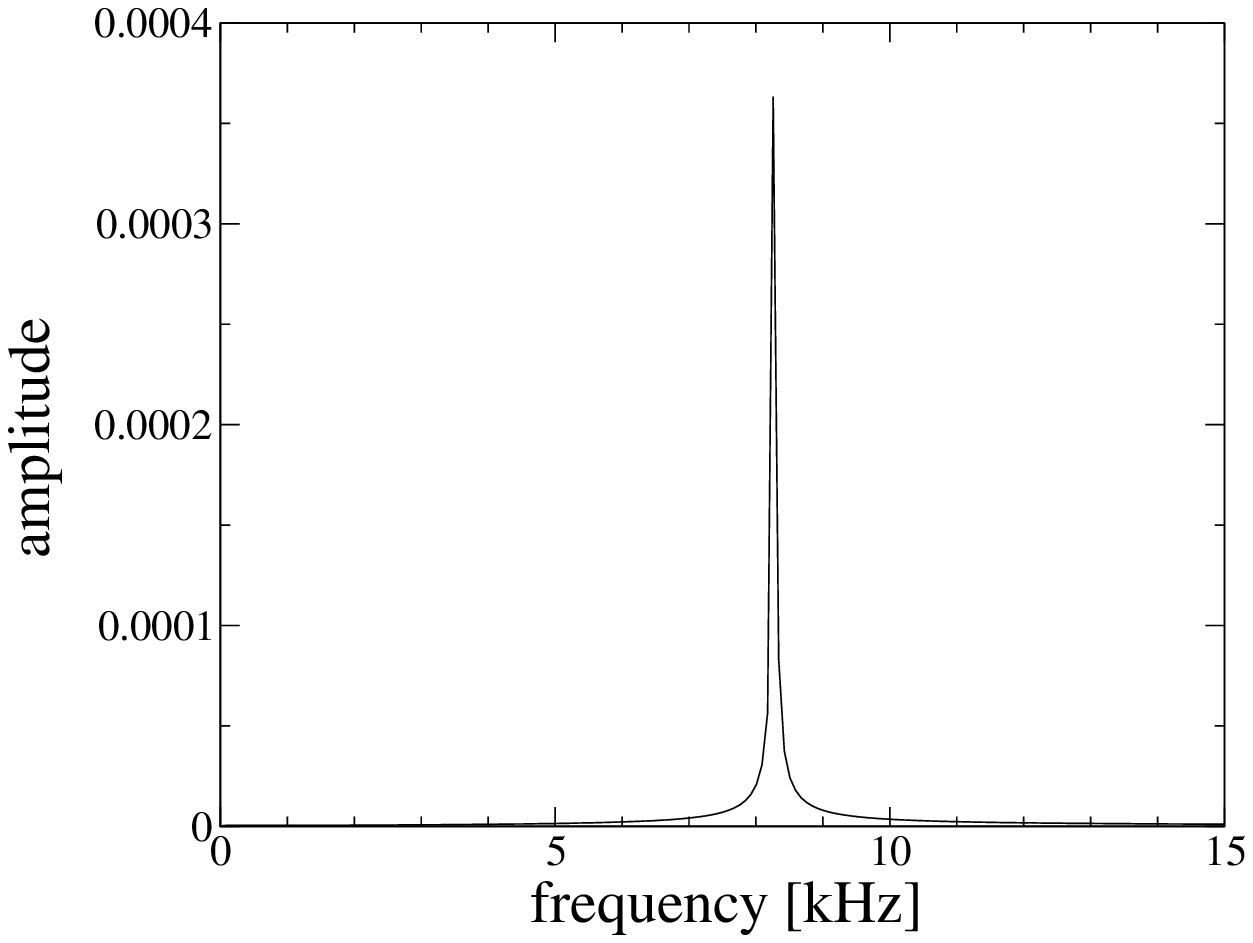}
 \caption{\label{fig:mode1}
 First overtone with maximal initial velocity $v(r_0,0)=20\mathrm{\frac{km}{s}}$ as in Fig.~\ref{fig:mode0}.}
\end{figure}

As numerical examples we consider the fundamental mode and the first overtone. The resulting radius functions $R(t)$ are shown in Figs.~\ref{fig:mode0}a and \ref{fig:mode1}a. As expected, $R(t)$ seems indeed to be a purely harmonic function of time. This can be illustrated more explicitly by looking at the corresponding Fourier spectrum as obtained with a discrete Fourier transform, see Figs.~\ref{fig:mode0}b and \ref{fig:mode1}b. Normally, in such a numerical simulation, one might expect to see a large peak at the corresponding eigenfrequency together with many superposed peaks (numerical noise) due to errors introduced by the numerical method. Here, however, we find nicely ``clean'' spectra, indicating that we have almost pure sinusoidal oscillations.

\begin{figure}\centering
 \includegraphics[scale=0.628]{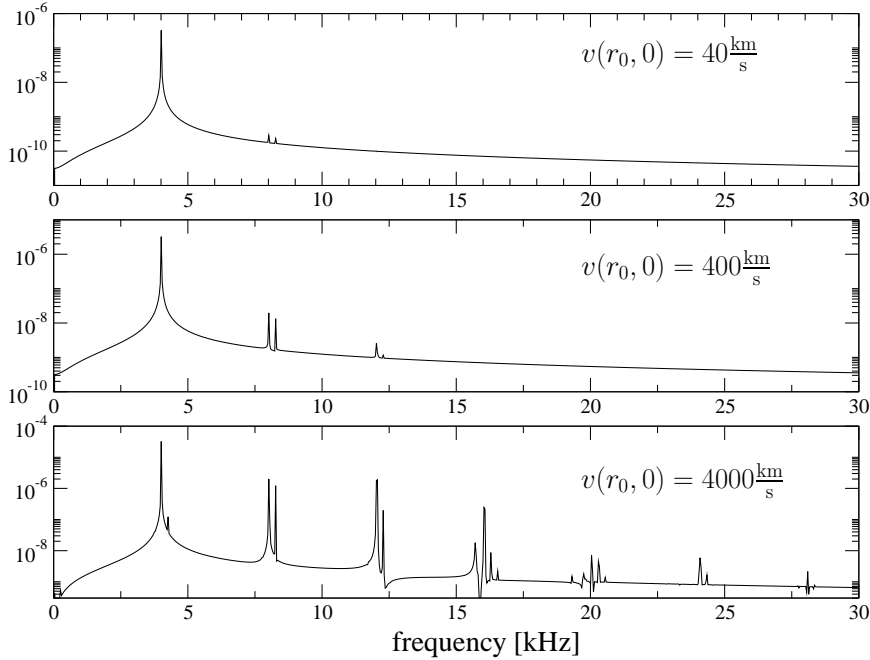}
 \caption{\label{fig:mode0-nonlin}
 Starting from data corresponding to the fundamental mode, also other modes are excited due to nonlinear effects. The diagrams show the spectra, obtained from Fourier transformation of the central mass density $\rho(0,t)$, for three simulations with different amplitudes of the initial velocity field.}
\end{figure}

Having discussed the linear regime, it is interesting to study larger deviations from equilibrium. In that case, nonlinear effects are not negligible anymore. An interesting nonlinear effect is the excitation of modes that are not present in the initial data. This can be demonstrated by starting from initial data corresponding to the fundamental mode, but with increasingly large amplitudes. For small amplitudes, we expect to find again the harmonic oscillations with the fundamental frequency. However, for higher amplitudes, we should  observe that other peaks appear in the spectrum (whose frequencies correspond to higher overtones or to linear combinations of several eigenfrequencies). And indeed, this effect shows clearly in the numerical examples, see Fig.~\ref{fig:mode0-nonlin}.

\mbox{}\pagebreak
\section{Cosmological models\label{sec:gowdy}}

As a second application of the fully pseudospectral scheme we study a particular cosmological model, namely \emph{Gowdy spacetimes} with spatial $S^3$ topology.

\subsection{The physical model}

Gowdy spacetimes \cite{Gowdy74} are relatively simple, but nontrivial inhomogeneous cosmological models within Albert Einstein's theory of general relativity. Note that the term ``cosmological model'' does not imply that we try to find an accurate description of our real universe. Instead it refers to a model in mathematical relativity that is assumed to reflect only \emph{some} aspects of the cosmos we live in. A major motivation for the study of these spacetimes comes from the desire to understand the mathematical and physical properties of cosmological singularities. 

Mathematically, Gowdy spacetimes are four-dimensional spacetimes characterized by existence of two spacelike and commuting Killing vectors (i.e.~by existence of two symmetries)\footnote{In addition, the precise definition of Gowdy spacetimes also includes that the so-called twist constants are zero.}. This allows us to introduce adapted coordinates in which everything effectively depends only on \emph{two} (instead of four) coordinates $\theta$ and $t$. Hence, we again arrive at a $1+1$ dimensional problem. Moreover, we consider \emph{vacuum} solutions to Einstein's field equations (i.e.~no matter and no electromagnetic or other fields are present) with vanishing cosmological constant. Under these and a few further assumptions\footnote{We assume compact, connected, orientable and smooth three manifolds.}, it turns out that the  spatial topology (the topology of slices $t=\textrm{constant}$) must be either $T^3$, $S^3$,  $S^2\times S^1$ or that of the lens space $L(p,q)$, cf.~\cite{Gowdy74} and  
\cite{Mostert,Neumann,Fischer}. Here, we restrict ourselves to the case of $S^3$ topology.

In order to obtain Gowdy solutions, we have to solve Einstein's field equations. In our setting, the essential part of these equations turns out to be equivalent to a single, complex equation, the \emph{Ernst equation} \cite{Ernst,BeyerHennig}
\begin{equation}\label{eq:EE}
 \Re(\E)\left(-\E_{,tt}-\cot t\,\E_{,t}
        +\E_{,\theta\theta}+\cot\theta\,\E_{,\theta}\right)
 =-(\E_{,t})^2+(\E_{,\theta})^2
\end{equation}
for the complex \emph{Ernst potential} $\E=\E(\theta,t)=:f(\theta,t)+\ii b(\theta,t)$. This potential contains basically all information about the geometry of the corresponding Gowdy spacetime.

An important question is whether $S^3$ Gowdy solutions that are regular at some time $t=t_0$ will also be regular in the past and future of $t_0$. An answer was provided in \cite{chrusciel1990}, where it was shown that regular data at some $t_0\in(0,\pi)$ guarantee regularity in the entire domain $t\in(0,\pi)$, $\theta\in(0,\pi)$. The boundaries $t=0$ and $t=\pi$ of this domain are expected to contain either curvature singularities or so-called \emph{Cauchy horizons}\footnote{A Cauchy horizon is a light-like boundary of the domain in which the Cauchy problem for the Einstein equations is valid. The region beyond a Cauchy horizon contains closed time-like geodesics, which implies a breakdown of causality and the possibility of time travel backwards in time.}. Furthermore, the boundaries $\theta=0$ and $\theta=\pi$ are regular up to coordinate singularities (axes singularities very much like the well-known axes singularities of spherical coordinates).

Here, we consider those Gowdy models with a  regular Cauchy horizon at the initial slice $t=0$. Properties of these spacetimes have been studied in \cite{BeyerHennig} and a family of exact solutions has been constructed in \cite{BeyerHennig2012}. The main result is that these Gowdy solutions automatically have a second Cauchy horizon at the final slice $t=\pi$, unless the initial Ernst potential satisfies one of the two conditions
\begin{equation}\label{eq:cond}
 b(\pi,0)-b(0,0)=\pm 4.
\end{equation}
In this case, the spacetime develops a curvature singularity at $t=\pi$, $\theta=0$ (for a `$+$' sign in the latter equation) or at $t=\pi$, $\theta=\pi$ (for a `$-$' sign), which leads to a divergent Ernst potential at the respective points.

In the following, we intend to solve the Ernst equation \eqref{eq:EE} numerically, starting from initial data at $t=0$. Note that, even though \eqref{eq:EE} is second order in time, we can only prescribe the initial values but not the time derivative of the Ernst potential. The reason is that \eqref{eq:EE} degenerates at the special hypersurface $t=0$ and leads to the regularity condition $\E_{,t}(\theta,0)=0$. Moreover, as a consequence of the $S^3$ topology and the fact that $t=0$ is a Cauchy horizon, the initial potential $\E(\theta,0) \equiv \E_0(\theta) \equiv f_0(\theta)+\ii b_0(\theta)$ must be a smooth function subject to the following conditions~\cite{BeyerHennig2012},
\begin{eqnarray}
 && f_0(0)=f_0(\pi)=0,\quad
 f_0(\theta)>0\textrm{ for }\theta\in(0,\pi),\\
 && f_{0,\theta}(0)=f_{0,\theta}(\pi)=0,\quad
 f_{0,\theta\theta}(0)=f_{0,\theta\theta}(\pi),\\
 && b_{0,\theta}(0)=b_{0,\theta}(\pi)=0,\quad
 b_{0,\theta\theta}(0)=b_{0,\theta\theta}(\pi)=2.
\end{eqnarray}

\subsection{Numerical details}

According to the above discussion, we expect to find a regular solution to the Ernst equation \eqref{eq:EE} with initial data at $t=0$ in the domain $(\theta,t)\in[0,\pi]\times[0,\pi)$, whereas the slice $t=\pi$ might contain singularities. Therefore, we intend to construct numerical solutions in the domain $[0,\pi]\times[0,t_\textrm{max}]$ for some $t_\textrm{max}\in(0,\pi)$. In the singular case, it is then particularly interesting to find out how close $t_\textrm{max}$ can be chosen to $t=\pi$.

The next step is to map this domain to unit squares. Here, we choose to represent the entire domain by a single square (in contrast to the larger number of time domains as used in the example of stellar pulsations). For that purpose, we might consider a simple linear rescaling of the coordinates $\theta$ and $t$. However, it turns out to be useful to choose $\cos\theta$ and $\cos t$ as new coordinates first and then to perform an appropriate rescaling, 
\begin{equation}
 \sigma=\frac12(1-\cos\theta),\quad
 \tau = \frac{1-\cos t}{1-\cos t_\textrm{max}}, 
\end{equation}
because this leads to a particularly nice form of the Ernst equation without trigonometric functions. 

It is then straightforward to decompose \eqref{eq:EE} into its real and imaginary parts to obtain a nonlinear system of two coupled equations for the two unknowns $f=\Re\E$ and $b=\Im\E$. Afterwards, the remaining steps of the fully pseudospectral scheme can be carried out without difficulty as described in Sec.~\ref{sec:method}. The only point that requires some careful attention is a degeneracy of the Ernst equation at the points $(0,0)$ and $(\pi,0)$, where the reformulation of \eqref{eq:EE} in terms of the spectral coordinates $\sigma$, $\tau$ turns out to be satisfied identically. However, one can derive additional non-trivial conditions at these points from the Ernst equation itself. See \cite{Hennig2009} for more details about the treatment of such ``exceptional points''.

\subsection{Numerical results}

We consider two numerical examples of solving the Ernst equation \eqref{eq:EE} with the fully pseudospectral scheme.

\begin{figure}\centering
 \includegraphics[scale=0.99]{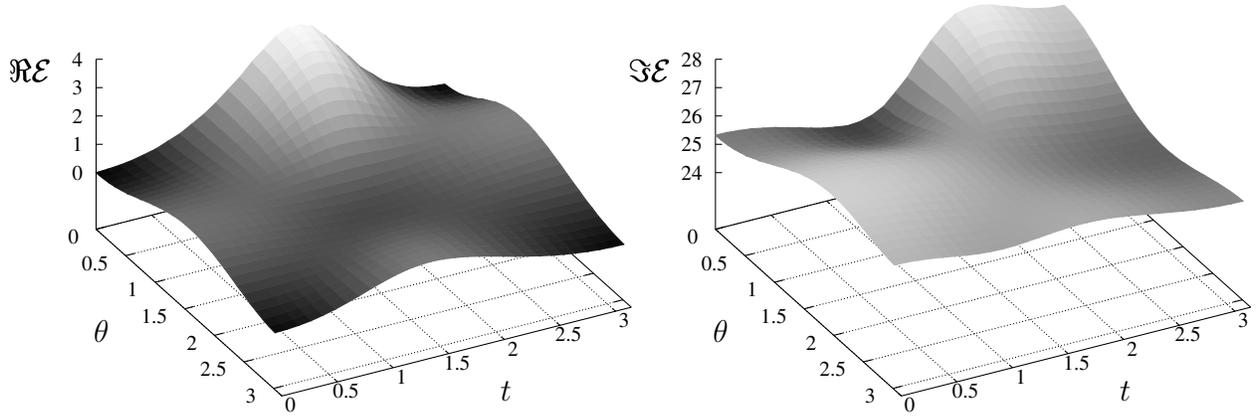}
 \caption{\label{fig:regular}
 Real and imaginary parts of the Ernst potential $\E=f+\ii b$ obtained from time evolution of the initial potential $f(\theta,0)=\sin^2\!\theta$, $b(\theta,0)=4(\sqrt{3}+3)\sqrt{2+\cos\theta}-2(\sqrt{3}+2)\cos\theta$. For these initial values we have $b(\pi,0)-b(0,0)=8-4\sqrt{3} \neq \pm 4$, which corresponds to a regular case without singularities.} 
\end{figure}

In the first example, we choose initial data that do \emph{not} satisfy the characteristic equation for singularities \eqref{eq:cond}. Then we know that the solution remains regular in the whole time interval $[0,\pi]$, and both of the boundaries $t=0$ and $t=\pi$ contain Cauchy horizons. The numerical result is illustrated in Fig.~\ref{fig:regular}. As expected, we obtain regular functions $\Re\E$ and $\Im\E$. 

\begin{figure}\centering
 \includegraphics[scale=0.99]{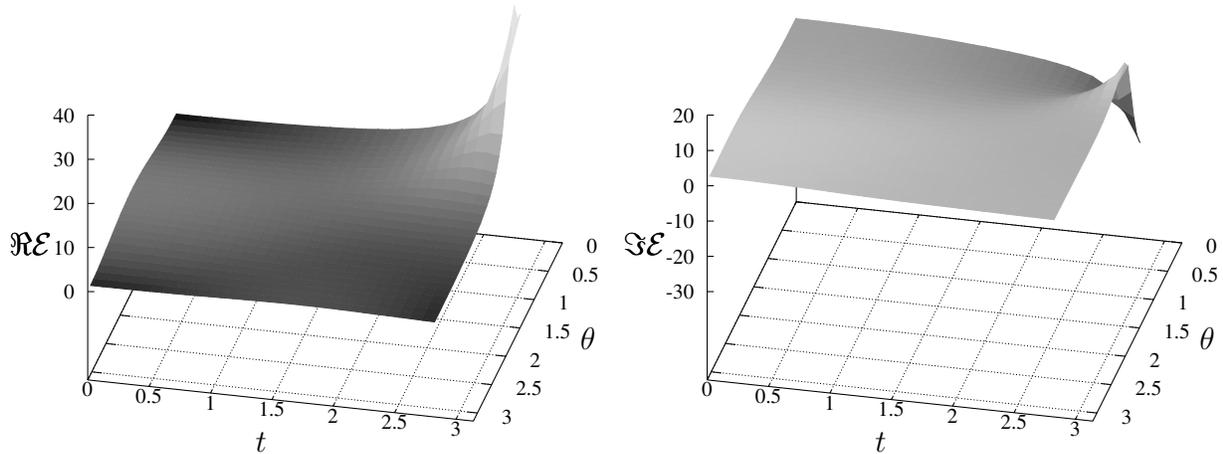}
 \caption{\label{fig:singular}
 Numerical values for the real and imaginary parts of the Ernst potential $\E=f+\ii b$ for the initial value problem with $f(\theta,0)=3\sin^2\!\theta$, $b(\theta,0)=\cos\theta\, (\cos^2\!\theta-\cos\theta-3)$. Here, we have $b(\pi,0)-b(0,0)=4$. As a consequence, the corresponding solution is singular and blows up at $t=\pi$, $\theta=0$. Interestingly, it is even possible to construct the \emph{exact} solution to this particular initial value problem for the Ernst equation, which is demonstrated in \cite{BeyerHennig2012}.}
\end{figure}

Before we study the accuracy of this result (see Sec.~\ref{sec:gowdycon}), we consider, as a second example, initial data subject to \eqref{eq:cond} with a plus sign on the right hand side. Consequently, we expect that the time evolution of these data leads to the formation of a singularity at $t=\pi$, $\theta=0$. The corresponding numerical solution, shown in Fig.~\ref{fig:singular}, indeed indicates  precisely this behaviour. 

\subsection{Convergence}\label{sec:gowdycon}

As in the case of stellar pulsation, a crucial point is the question of convergence and numerical accuracy. In order to answer this question in the present case of $S^3$ Gowdy spacetimes, we can make use of results from \cite{BeyerHennig}. In that article, explicit formulae for the Ernst potential at the boundaries $\theta=0$, $\theta=\pi$ and $t=\pi$ in terms of the initial potential at $t=0$ have been derived. In particular, for $\E$ at $\theta=0$ we have
\begin{equation}\label{eq:formula}
 \E(0,t)= \frac{\ii[b(0,0)-2(1-\cos t)]\E(t,0)+b^2(0,0)}
                        {\E(t,0)-\ii[b(0,0)+2(1-\cos t)]},
\end{equation}
where $\E(t,0)$ means $\E(\theta=t,t=0)$.
Using this formula, we can, for example, calculate the exact value of $f=\Re\E$ at $\theta=0$, $t=t_\textrm{max}$ and compare it with the numerical value. The corresponding relative error may then be used as a measure for the numerical accuracy.

In the following, we apply this method for studying the accuracy of the regular and the singular example evolutions presented in the previous subsection. In both cases, we consider a number of different spectral resolutions and maximal times $t_\textrm{max}$.

\begin{figure}\centering
 \hspace{.2cm}(a)\hspace{7.95cm}(b)\hfill\mbox{}\\
 \includegraphics[scale=0.63]{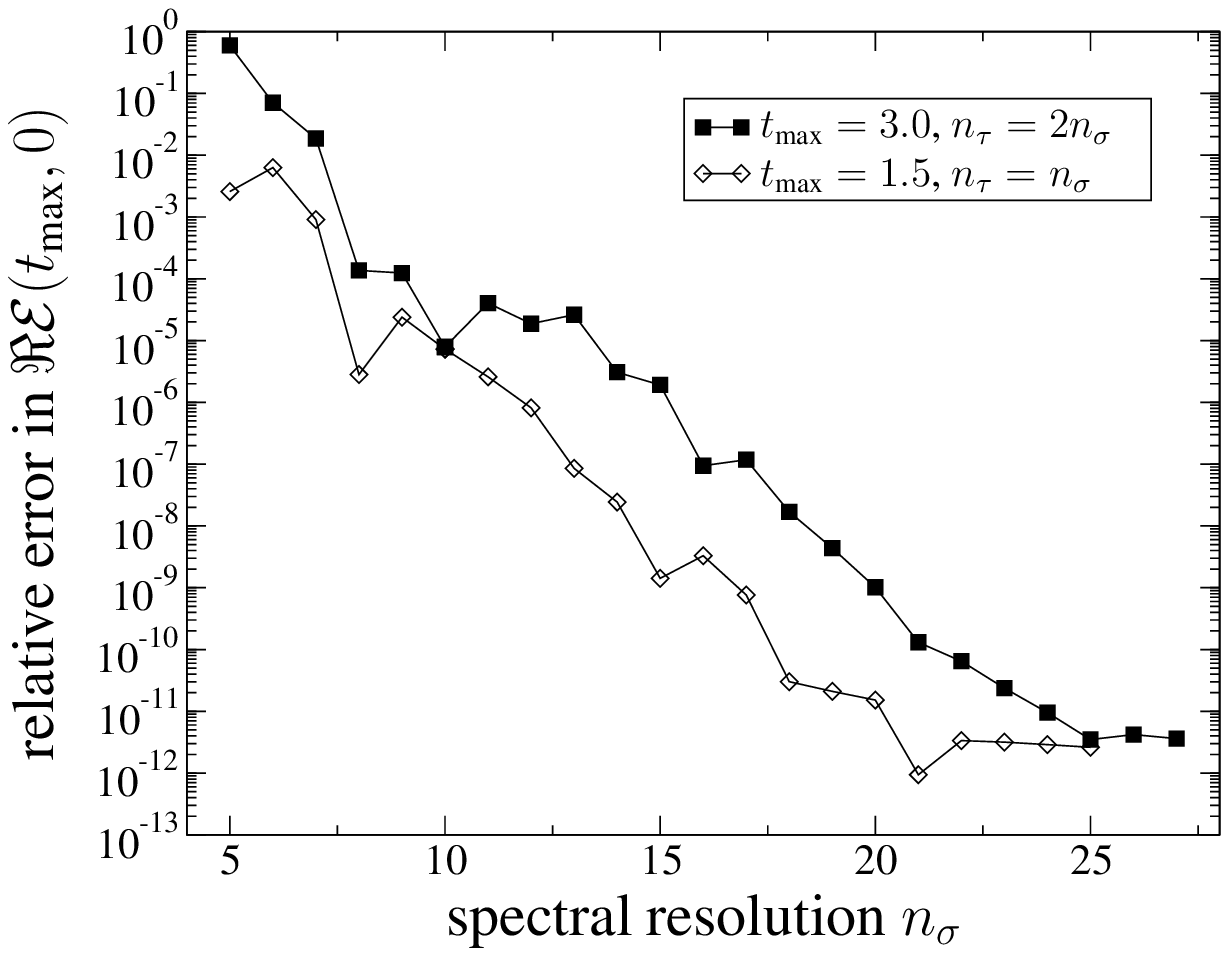}\quad\,
 \includegraphics[scale=0.63]{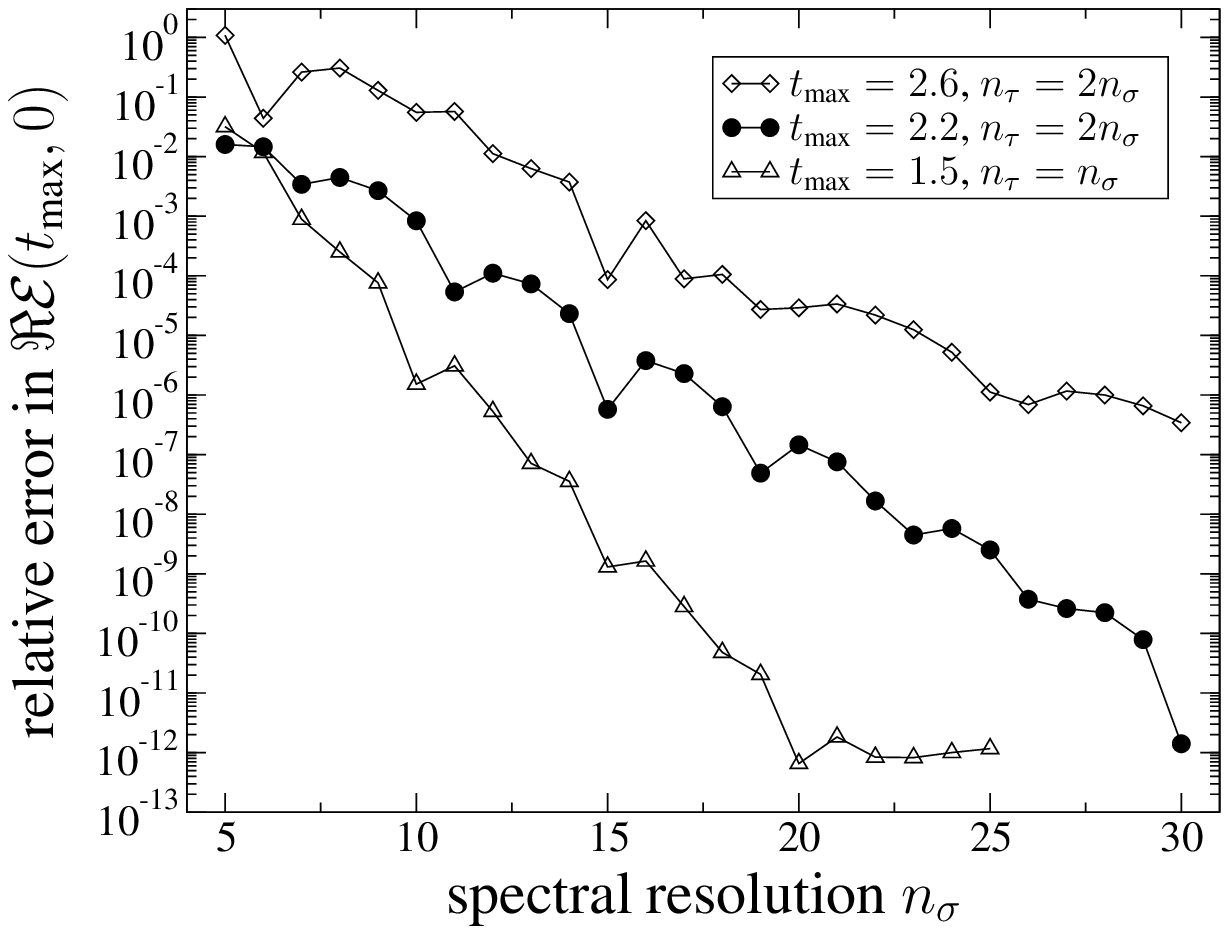}
 \caption{\label{fig:GowdyCon}
 Test of the numerical accuracy: Two sets of initial data are evolved for different choices of the maximal time $t_\textrm{max}$. Afterwards, the numerical value for the real part of the Ernst potential at $t=t_\textrm{max}$, $\theta=0$ can be compared with the exact value from the analytical formula \eqref{eq:formula}. The diagrams show the relative error in the numerical value.\\
 (a) Initial data as in Fig.~\ref{fig:regular}, corresponding to a solution with a regular Cauchy horizon at $t=\pi$.\\
 (b) Initial data as in Fig.~\ref{fig:singular}, leading to a singularity at $t=\pi$, $\theta=0$.}
\end{figure}

The accuracy for the regular example is shown in Fig.~\ref{fig:GowdyCon}a. We see that the error decreases with increasing resolution until saturation between $10^{-11}$ and $10^{-12}$ is reached. Hence, the numerical solution in the regular case is highly accurate. By comparing the error curve for $t_\textrm{max}=1.5$ with the curve for $t_\textrm{max}=3.0$, we see that, for larger $t_\textrm{max}$, a larger resolution is required in order to reach the saturation level. But this is clear since evidently more Chebyshev polynomials are required if a function is to be accurately approximated on a larger domain.

We can do the same investigation for the singular example studied in the previous subsection. Since the corresponding Ernst potential diverges at $t=\pi$, $\theta=0$, we cannot expect very accurate solutions if $t_\textrm{max}$ approaches $\pi$. Indeed, this expected drop in the accuracy can clearly be seen in  Fig.~\ref{fig:GowdyCon}b. For $t_\textrm{max}=1.5$, the error decreases to about $10^{-12}$ for $n_\sigma=20$ (and $n_\tau=20$). In order to reach the same level of accuracy for $t_\textrm{max}=2.2$, we already need a resolution of $n_\sigma=30$ (and $n_\tau=60$). Finally, for $t_\textrm{max}=2.6$ the error has ``only'' decreased to about $10^{-6}$ for the maximal considered resolution of $n_\sigma=30$ (and $n_\tau=60$). This is still very accurate compared to errors in typical low-order finite difference algorithms, but definitely much below the ``usual spectral accuracy'' (which is close to machine accuracy). Clearly, a huge number of Chebyshev polynomials is needed to approximate functions in 
the vicinity of a singularity. 

A possible approach for increasing the numerical accuracy near singularities could be to split the time domain $[0,t_\textrm{max}]$ into several subdomains (as done in the case of stellar pulsations). Then, in the first domains a low resolution $n_\tau$ would be sufficient, whereas larger resolutions could be used in the domains close to the singularity. In this way, one could ``concentrate'' higher resolutions where they are needed. Another possibility would be to rescale the coordinates in such a way that steep gradients near the singularity are reduced. An example for this procedure can be found on p.~119 of \cite{Meinel}.

Finally, we note that the error curves in the logarithmic plots in Fig.~\ref{fig:GowdyCon} are more or less linear, i.e.~we again observe \emph{exponential convergence} (with higher convergence rates for regular solutions and lower convergence rates in singular cases). 

\section{Discussion\label{sec:discussion}}

We have studied the applicability of the fully pseudospectral numerical scheme, introduced in \cite{Hennig2009}, to solving time dependent physical problems. In particular, we have considered stellar pulsations of spherically symmetric, polytropic Newtonian stars and the time evolution of Gowdy spacetimes with spatial $S^3$ topology.

The basic idea of the method is to solve initial-boundary value problems for nonlinear PDEs with Chebyshev expansions in spatial \emph{and} time directions. As a consequence, space and time are treated relatively equal. Moreover, we obtain a highly implicit method for which the numerical solution in entire spacetime domains is obtained simultaneously.

Some details of the numerical implementation were different in our two examples (free-boundary problem and several time domains for stellar oscillations, but fixed boundaries and only a single time domain for Gowdy spacetimes). However, the final outcome was the same and confirmed the observations from the examples of the simple scalar waves equation discussed in \cite{Hennig2009}: with a fully pseudospectral scheme, we obtain highly accurate numerical solutions with relative error in the order $10^{-11}$ to $10^{-13}$ (corresponding to up to $13$ significant digits, which is close to the machine accuracy of our double precision code). This is possible with moderate spectral resolutions, provided the solutions are sufficiently regular. On the other hand, close to singularities, much higher resolutions are required for highly accurate solutions; otherwise, the accuracy can drop significantly. However, both in the regular and singular cases, the method possesses a geometric convergence rate and the error 
decreases exponentially with the spectral resolution. Moreover, the example of stellar pulsations has demonstrated that stable long-term evolutions are possible, namely evolutions over many oscillation periods.

These results should encourage the application of the fully pseudospectral scheme to wide classes of problems in physics and other areas. The high accuracy of the method will allow for very precise numerical studies, very clean spectra, accurate preservation of conserved quantities, etc. 

In the future, it would be desirable to develop the method further in order to allow for a treatment of higher dimensional problems (e.g.~time evolution of axisymmetric rather than spherically symmetric stars). An important step in this direction would be to replace the computationally most expensive part of the method, namely the matrix inversion in the Newton Raphson scheme, by an iterative method\footnote{The complexity of the matrix inversion via LU-decomposition as used here is proportional to the third power of the matrix dimension, and the matrix dimension itself is proportional to $n_\sigma\times n_\tau$. Moreover, in the examples discussed here, about two to four matrix inversions are required in each spectral domain. Hence, for $n_\sigma\approx n_\tau=n$, the complexity of the fully pseudospectral scheme is $\mathcal O(n^6)$. As an example, on my PC, the inversion of the matrix appearing in the calculation of the Newtonian star takes about $0.18$\,s for $n=15$ and about $7.8$\,s for $n=26$. For 
comparison, a low-order finite difference algorithm would only have a complexity of $\mathcal O(n^2)$, but, on the other hand, require a huge number of gridpoints to achieve an accuracy comparable to ``spectral accuracy''. Since a small $n$ is sufficient in the pseudospectral scheme, the direct matrix inversion is reasonable in $1+1$ dimensions, whereas for $2+1$ or higher dimensions an iterative inversion method will be essential.}.

\begin{acknowledgments}
 I am indebted to Marcus Ansorg, who had introduced me to the topic of spectral methods during our delightful collaboration at the Max Planck Institute for Gravitational Physics, Potsdam. Moreover, I would like to thank Marcus Ansorg and J\'er\^ome Novak for many valuable discussions, Ernazar Abdikamalov for providing me with data from a finite difference simulation and Ben Whale for commenting on the manuscript. 
\end{acknowledgments}


\end{document}